\documentstyle[aps,prl]{revtex}
\begin{document}
\twocolumn[{
\title{Fusion Rules in Turbulent Systems with Flux Equilibrium}
  \author {Victor L'vov$^\ast$ and Itamar  Procaccia$^\dagger$
}
\address{Departments of~~$^\ast$Physics of Complex Systems
  {\rm and}~~$^\dagger$Chemical Physics,\\
   The Weizmann Institute of Science, Rehovot 76100, Israel
 }
 \maketitle
\widetext
 \begin{abstract}
 \leftskip 54.8pt
 \rightskip 54.8pt
Fusion rules in turbulence specify the analytic structure of many-point
correlation functions of the turbulent field when a group of coordinates
coalesce.  We show that the existence of flux equilibrium in
fully developed turbulent systems combined with a direct cascade induces
universal
fusion rules. In certain examples these fusion rules suffice to compute
the multiscaling exponents exactly, and in other examples they give rise
to an infinite number of scaling relations that constrain enormously
the structure of the allowed theory.
\end{abstract}
 \leftskip 54.8pt
}]
\narrowtext
In a series of recent papers elements of the analytic theory of
Navier-Stokes turbulence \cite{95LP-a,95LP-b,95LP-c,95LP-d} and
passive-scalar turbulent advection \cite{94Kra,94LPF,95KYC,95FGLP} were
presented. In this Letter we explain that the structure of the essential
part of these theories is economically summarized by a set of ``fusion
rules" that determine the analytic structure of $n$-point correlation
functions when a group of coordinates coalesces. We show here that the
fusion rules can be deduced from very few general assumptions about the
nature of the universal flux equilibrium that exists in fully developed
turbulent systems. Of course, the same fusion rules can be also
established by direct calculations in specific examples.  We first deduce
the fusion rules, then we exemplify their utility in determining scaling
exponents, and lastly we demonstrate how in one example the fusion rules
follow from first principles.

Consider a turbulent field ${\bf u}({\bf r},t)$ which is either a vector or a
scalar and denote the difference $ {\bf w}({\bf r}_1|{\bf r}_2,t)\equiv
{\bf u}({\bf r}_2,t)- {\bf u}({\bf r}_1,t)$. We discuss the statistical
properties of
the turbulent field in terms of simultaneous many-point generalized structure
functions of the types
\FL
\begin{eqnarray}
&&\!\!\!\!\!\!\!
  {\cal S}_n({\bf r}_0|{\bf r}_1,{\bf r}_2\dots{\bf r}_n) \!=\!
\left<{\bf w}({\bf r}_0|{\bf r}_1)\,
{\bf w}({\bf r}_0|{\bf r}_2)\dots {\bf w}({\bf r}_0|{\bf r}_n)\right>,
\label{Sn} \\
&&  \!\!\!\!\!\!
{\cal S}_{n,m}({\bf r}_0,{\bf r}'_0|{\bf r}_1\dots{\bf r}_n;{\bf r}_{n+1}\dots
{\bf r}_{n+m})
=\big\langle{\bf w}({\bf r}_0|{\bf r}_1)
\label{Snm}\\
&\times&
{\bf w}({\bf r}_0|{\bf r}_2)\dots
{\bf w}({\bf r}_0|{\bf r}_n) {\bf w}({\bf r}'_0|{\bf r}_{n+1})\dots
{\bf w}({\bf r}'_0|{\bf r}_{n+m})\big\rangle \,,
\nonumber
\end{eqnarray}
etc. Note that when ${\bf u}$ is a vector the $n$-point correlation is an
$n$-rank tensor. In other words these are correlation functions of
differences with respect to one, two or more reference points. The
class of systems that we discuss are driven on a characteristic scale
referring as the outer scale $L$.  This driving can be achieved by
either a time dependent low frequency ``stirring force" or by specifying
given values of ${\bf u}$ at a set of ``boundary" points with a
characteristic separation $L$ away from our observation points
${\bf r}_0,{\bf r}'_0$, etc.  The systems have dissipation (viscosity,
diffusivity etc.) and in the dissipationless limit there exists an integral
of motion which we refer to as ``energy".  We consider systems with a
``direct" cascade in which the intake of energy on the large scales is
balanced by dissipation on the characteristic small scale $\eta$. Fully
developed turbulence is associated with the limit $L/\eta \to \infty$.

We invoke two assumptions of the Kolmogorov \cite{41Kol} type:

\noindent
{\bf 1.} Universality of the fine scale structure of turbulence; by this
we mean that the correlation functions of the type (\ref{Sn}), (\ref{Snm})
have a universal functional dependence on their arguments as long as all
the separation distances involved are much smaller than $L$. This means
that we can fix an arbitrary set of velocity differences on the scale of $L$,
and the correlation functions will depend on their precise choice
only via an overall factor that depends on the $L$-scale motions.
Mathematically this is expressed as the following property of the
conditional average:
\begin{eqnarray}
&&\!\!\!\!\!\!\!\!\!\!
\big\langle {\bf w}({\bf r}_0|{\bf r}_1)
{\bf w}({\bf r}_0|{\bf r}_2)\dots {\bf w}({\bf r}_0|{\bf r}_n) \Big |{\bf
w}({\bf r}_0|{\bf R}_1)
{\bf w}({\bf r}_0|{\bf R}_2)
\label{assum1} \\
&\dots&{\bf w}({\bf r}_0|{\bf R}_N) \big\rangle
= {\cal S}_n({\bf r}_0|{\bf r}_1,{\bf r}_2\dots{\bf r}_n) \Phi_{n,N}
({\bf r}_0|{\bf R}_1\dots {\bf R}_N)
\nonumber
\end{eqnarray}
for $|{\bf R}_i -{\bf r}_0| \sim L$ and $|{\bf r}_i -{\bf r}_0| \ll L$.

\noindent
{\bf 2.} Scale invariance: all the correlation functions are
homogeneous functions of their arguments in the core of the inertial
interval $\eta \ll |{\bf r}_i -{\bf r}_0| \ll L$:
\begin{equation}
{\cal S}_n(\lambda{\bf r}_0 |\lambda{\bf r}_1,\lambda{\bf r}_2\dots\lambda{\bf
r}_n) =
\lambda^{\zeta_n}{\cal S}_n({\bf r}_0|{\bf r}_1,{\bf r}_2\dots{\bf r}_n),
\label{assum2}
 \end{equation}
where $\zeta_n$ is the scaling exponent of the $n$-order structure
function. We are particularly interested in systems in which $\zeta_n$ is
a nonlinear function of $n$. We refer to such systems as ``multiscaling".

The derivation of these two properties from first principles
differs from system to system. In this Letter we discuss the fusion rules
and their consequences in systems for which these assumptions are valid.
The first set of fusion rules that we derive concerns ${\cal S}_n$ when $p$
points $(p<n)$ coalesce with ${\bf r}_0$, (so that the typical separation
from ${\bf r}_0$ is $r$) and all the other separations remain much larger, of
the order of $R$, $r\ll R \ll L$.  Without loss of generality we can
choose these $p$ coordinates as ${\bf r}_1,{\bf r}_2,\dots {\bf r}_p$.  We
claim
that
\begin{eqnarray}
&& {\cal S}_n({\bf r}_0|{\bf r}_1,{\bf r}_2,\dots,{\bf r}_n)
\nonumber \\
&=&
{\cal S}_p({\bf r}_0|{\bf r}_1,{\bf r}_2,\dots,{\bf r}_p)
\Psi_{n,p}({\bf r}_0|{\bf r}_{p+1},{\bf r}_{p+2},\dots,{\bf r}_n) \ ,
\label{fusion1}
\end{eqnarray}
where $\Psi_{n,p}({\bf r}_0|{\bf r}_{p+1},{\bf r}_{p+2},\dots,{\bf r}_n)$ is a
homogeneous function with a scaling exponent $\zeta_n-\zeta_p$. The
derivation of the fusion rule (\ref{fusion1}) follows from Bayes' theorem
and assumptions 1,2. We write
\FL
\begin{eqnarray}
&& {\cal S} _n({\bf r}_0|{\bf r}_1,{\bf r}_2,\dots,{\bf r}_n)=\int
d{\bf w}({\bf r}_0|{\bf r}_{p+1})\dots d{\bf w}({\bf r}_0|{\bf r}_n)
\nonumber\\
&\times&
{\bf w}({\bf r}_0|{\bf r}_{p+1})\dots {\bf w}({\bf r}_0|{\bf r}_n)
{\cal P}[{\bf w}({\bf r}_0|{\bf r}_{p+1}\dots {\bf w}({\bf r}_0|{\bf r}_n)]
\nonumber \\
&\times&\big\langle {\bf w}({\bf r}_0|{\bf r}_1),
{\bf w}({\bf r}_0|{\bf r}_2)\dots {\bf w}({\bf r}_0|{\bf r}_p)\Big |{\bf
w}({\bf r}_0|{\bf r}_{p+1})
\nonumber \\
&\times&
{\bf w}({\bf r}_0|{\bf r}_{p+2})\dots {\bf w}({\bf r}_0|{\bf r}_n) \big\rangle
\ ,
\label{bayes}
\end{eqnarray}
where ${\cal P}[{\bf w}({\bf r}_0|{\bf r}_{p+1}\dots {\bf w}({\bf r}_0|{\bf
r}_n)]$ is the
probability to see the tensor ${\bf w}({\bf r}_0|{\bf r}_{p+1}\dots
{\bf w}({\bf r}_0|{\bf r}_n)$.  Next comes the central idea of this Letter: the
properties of flux equilibrium and the universal structure of the
correlation functions on the scale $r$ are the same independen of whether  we
force the system on the scale $L\gg r$ or on the scale $R\gg r$.  The
conditional average in (\ref{bayes}) is proportional to ${\cal S}_p$, and
hence (\ref{fusion1}). This result seems rather obvious at this point, but
we will see that it leads to a totally unconventional scaling structure of
the theory. We should stress that for Navier-Stokes and passive scalar
advection
these fusion rules for $p=2$ were derived from first principles
\cite{95LP-d,95FGLP}.

The next set of fusion rules is obtained for the structure function ${\cal
S}_{n,m}$ of (\ref{Snm}) when two groups of $p\le n$ and $q\le m$ points
coalesce onto ${\bf r}_0$ and ${\bf r}'_0$ respectively. The separation
between the groups of point is large and of the order of $R$. The
derivation of the fusion rules is now obvious, with the result
\begin{eqnarray}
&& {\cal S}_{n,m}({\bf r}_0,{\bf r}'_0|{\bf r}_1,\dots,{\bf r}_n;{\bf
r}_{n+1},\dots ,
{\bf r}_{n+m})
\nonumber  \\
&=&{\cal S}_p({\bf r}_0|{\bf r}_1,{\bf r}_2,\dots,{\bf r}_p)
{\cal S}_q({\bf r}_0|{\bf r}_1,{\bf r}_2,\dots,{\bf r}_q)
\nonumber\\
&\times&  \Psi_{n,m,p,q}({\bf r}_0,{\bf r}'_0|{\bf r}_{p+1},
\dots,{\bf r}_n; {\bf r}_{n+q+1},\dots,{\bf r}_{n+m}) \ .
\label{fusion2}
\end{eqnarray}
The scaling exponent of $\Psi_{n,m,p,q}$ is $\zeta_{n+m}-\zeta_p-\zeta_q$.
Note that the fusion rules (\ref{fusion1}) and (\ref{fusion2}) are {\it
not} decompositions into products of lower order correlation functions,
and the functions $\Psi$ are not correlations of velocity differences
across large separations. In fact we will show that $\Psi$
is much larger than the corresponding correlation functions in all
situations with multiscaling.  Evidently one can derive similar
fusion rules for three, four or more groups of coalescing points with
large separations between the groups. The structure of the resulting
correlation function will be a product of the correlation function
associated with each group times some function $\Psi$ of big separations
which carries the overall exponent.

Next we discuss fusion rules for correlation functions that include gradient
fields. These rules depend on the type of rotational invariant that one
can define from the tensors that appear after taking gradients. We will only
consider the lowest order invariant which is a scalar under rotation. For
passive scalars $T$ this is $|\nabla T\cdot \nabla T|^2$ and for a vector ${\bf
u}$
the quantity $|\nabla{\bf u}|^2$ is the square of the strain tensor $
s_{ij}s_{ij}$
where
$s_{ij}\equiv (\partial u_i/\partial r_j+\partial u_j/\partial r_i)/2$.
Consider the quantity
\begin{eqnarray}
J_{2p,n}({\bf r}_0|{\bf r}_{2p+1}\dots{\bf r}_n)&=&\langle|\nabla{\bf u}({\bf
r}_0)|^{2p}
\nonumber \\
&\times& {\bf w}({\bf r}_0|{\bf r}_{2p+1})\dots{\bf w}({\bf r}_0|{\bf
r}_n)\rangle \ . \label{Jpn}
\end{eqnarray}
To evaluate $J_{2p,n}$ we consider a related object in which all the
gradients are taken at different points:
\begin{eqnarray}
&&\tilde J_{2p,n}= \nabla^{i_1}_{r_1}\nabla^{j_1}_{r'_1} \nabla^{i_2}_{r_2}\
\nabla^{j_2}_{r'_2}\dots \nabla^{i_p}_{r_p}\nabla^{j_p}_{r'_p}
{\cal C}^{i_1,j_1\dots i_p,j_p}_{k_1,l_1\dots k_p,j_p}
\label{tricky}\\
&\times& \langle w^{k_1}({\bf r}_0|{\bf r}_1)w^{l_1}({\bf r}_0|{\bf r}'_1)
\dots w^{k_p}({\bf r}_0|{\bf r}_p)w^{l_p}({\bf r}_0|{\bf r}'_p) \nonumber \\
&\times&{\bf w}({\bf r}_0|{\bf r}_{2p+1})  \dots {\bf w}({\bf r}_0|{\bf r}_n)
\rangle ,\nonumber
\end{eqnarray}
where the contraction tensor ${\cal C}$ ensures that the required scalar is
obtained. We will represent this quantity in a compact form without displaying
all the tensor indices as
$\tilde J_{2p,n}({\bf r}_0|{\bf r}_1,{\bf r}_2\dots{\bf r}_n)=
\nabla_{r_1} \nabla_{r'_1} \dots \nabla_{r'_p} {\cal C}{\cal S}_n({\bf r}_0|
{\bf r}_1,{\bf r}'_1\dots{\bf r}_n)$.
The quantity (\ref{tricky}) gives us $J_{2p,n}$ when the $2p$ first points
coalesce together with ${\bf r}_0$, whereas all the rest of the coordinates
remain
a typical distance $R$ from ${\bf r}_0$. When $R$ is in the inertial interval
we
expect scaling behaviour in terms of $R$,
\begin{equation}
J_{2p,n}\propto R^{-\xi(2p,n)} \ . \label{xi}
\end{equation}
Considering the distances between all the coalescing points to be in the
inertial range we apply (\ref{fusion1}) and find for $2p$ coalescing points
\begin{eqnarray}
&&\!\!\!\!\!\! J_{2p,n}({\bf r}_0|{\bf r}_1,{\bf r}_2\dots{\bf r}_n)=
\nabla_{r_1} \dots \nabla_{r'_p} {\cal S}_{2p}({\bf r}_0|{\bf r}_1,{\bf
r}'_1\dots
{\bf r}'_p) \nonumber\\
&\times&
\Psi_{n,2p}({\bf r}_0|{\bf r}_{2p+1},{\bf r}_{2p+2},\dots,{\bf r}_n)\ .
\label{Jpnfused}
\end{eqnarray}
We expect that $J_{2p,n}$ is independent of the first $2p$ coordinates when
the distances between them are well in the viscous regime.  Our next
fundamental assumption is that there exists a characteristic viscous
length $\eta(2p,n,R)$ at which $J_{2p,n}$ crosses smoothly from inertial
range behaviour to dissipative behaviour with respect to the first $2p$
coordinates.  This allows us to evaluate the coalescing gradients by
taking the $2p$ separations to be $\eta(2p,n,R)$:
\begin{eqnarray}
&&\!\!\!\!\!\!\!\!  J_{2p,n}({\bf r}_0|{\bf r}_1,{\bf r}_2,\dots,{\bf r}_n)\sim
\eta(2p,n,R)^{\zeta_{2p}-2p}
\label{Jpneta}\\
&\times&
\Psi_{n,2p}({\bf r}_0|{\bf r}_{2p+1},{\bf r}_{2p+2},\dots,{\bf r}_n)
\quad (2p{\rm~coalescing~points}).\nonumber
\end{eqnarray}
If there are two groups of coalescing points with gradients, with $p$
points coalescing onto ${\bf r}_0$ and $q$ points coalescing on ${\bf r}'_0$
respectively, we consider $J_{p,q,n,m}$ (where as before $n+m\geq p+q$ is
the total number of points). The rule for $p$ and $q$ coalescing points
is
\begin{eqnarray}
&&\!\!\!\!\!\!\!\! J_{p,q,n,m}({\bf r}_0|{\bf r}_1,{\bf r}_2,\dots,{\bf
r}_n)\sim
\eta(p,n,R)^{\zeta_p-p} \eta(q,n,R)^{\zeta_q-q}
\nonumber\\
&\times&\Psi_{n,m,p,q}({\bf r}_0|{\bf r}_{p+1},{\bf r}_{p+2},\dots,{\bf r}_n)
\ .
\label{Jpneta2} \
\end{eqnarray}
The generalization of this fusion rule for three or more groups of
coalescing points with gradients is obvious.

This is as much as one can do in general. Now the crucial issue is how
$\eta(2p,n,R)$ depends on its arguments. The simplest version of the theory
comes about when the dissipative length is independent of $R$,
$\eta(2p,n,R)=\eta(2p,n)$.  This is realized for example in passive scalar
convection as is shown below.  In this case the fusion rules imply various
sets of scaling relations. For example the exponents $\xi(2p,n)$ of
$J_{2p,n}$ are given by
\begin{equation}
\xi(2p,n) =\zeta_n-\zeta_{2p} \ . \label{scale}
\end{equation}
As another example of scaling relations consider the correlation functions
\begin{equation}
K^{(2s)}_{\epsilon\epsilon}(R)\equiv
\left<|\nabla {\bf u}({\bf r})|^{2s}|\nabla {\bf u}({\bf r}+{\bf
R})|^{2s}\right>\propto
R^{-\mu(2s)} \ . \label{Ks}
\end{equation}
{}From (\ref{Jpneta2}) in the case $n=m=p=q=2s$ we get a set of scaling
relations
\begin{equation}
\mu(2s)=2\zeta_{2s}-\zeta_{4s} \ . \label{mus}
\end{equation}
Next we can consider a correlation of $l$ gradient fields with the same
power, (i.e. $|\nabla {\bf u}|^{2s}$) at $l$ different points which are
separated by a distance of the order of $R$.  The corresponding exponent
$\mu(l,2s)$ is
\begin{equation}
\mu(l,2s) = l \zeta_{2s}-\zeta_{2sl} \ . \label{muls}
\end{equation}
This algebra can be generalized to any correlation of powers of $|\nabla
{\bf u}|^2$. For example, the scaling exponent $\mu(p_1,p_2,\dots p_n)$ of a
correlation of fields $\langle |\nabla {\bf u}({\bf r}_1)|^{p_1}$$|\nabla
{\bf u}({\bf r}_2)|^{p_2} \dots |\nabla {\bf u}({\bf r}_n)|^{p_n}\rangle$ in
which all
the separations is of the order of $R$ is
\begin{equation}
\mu(p_1,p_2,\dots p_n) = \sum_{j=1}^n \zeta_{p_j} - \zeta_{\bar p}\ , \quad
\bar p={\sum p_j} \ .
\label{crazy}
\end{equation}
In usual operator algebras \cite{69Kad,69Wil,69Pol,84ZBP} every local
field is associated with a leading exponent and the correlation function
scales with the sum of these exponents. In this case the algebra is
different. There is a global exponent $\zeta_{\bar p}$ from which one
subtracts the sum of individual exponents $\zeta_{p_j}$. In multiscaling
situations the global exponent is a nonlinear function of $\bar p=\sum p_j$.
Accordingly it is not a property of every individual field. We note here
without demonstration that invariants of the gradient field tensors other
than scalars are associated with other individual exponents.

The range of applicability of these fusion rules should be understood on
the basis of the equations of motion for any given system. As an example
we explain here briefly why they are applicable for Kraichnan's model
\cite{68Kra} of passive scalar convection with a driving velocity field
that is $\delta$-correlated in time. It was shown \cite{68Kra} that the
cumulant part $F_{2n}^c$ of the $2n$-order correlation function $F_{2n}=
\left<T({\bf r}_1,t)T({\bf r}_2,t)\dots T({\bf r}_{2n},t)\right>$ satisfies for
$n>1$ the exact homogeneous differential equation
\begin{equation}
\Big [-\kappa\sum_{\alpha=1}^{2n}\nabla^2_\alpha
+ \hat {\cal B}_{2n}\Big ]
F_{2n}^c ({\bf r}_1,{\bf r}_2,\dots ,{\bf r}_{2n})=0 \ , \label{diffeq}
\end{equation}
where $\kappa$ is the molecular diffusivity and $\nabla^2_\alpha$ is the
Laplacian operator acting on ${\bf r}_\alpha$. The operator $\hat {\cal
B}_{2n}$ is the sum of the binary  operators $\hat {\cal B}_{\alpha
\beta}$:
\begin{equation}
\hat {\cal B}_{\alpha \beta} \equiv h_{i,j}({\bf r}_\alpha \!-\!{\bf r}_\beta)
{\partial^2
\over \partial r_{\alpha,i}\partial r_{\beta,j}}, \
\hat {\cal B}_{2n} =\!\! \sum_{\alpha> \beta=1}^{2n} \hat {\cal B}_{\alpha
\beta}.
\label{B}
\end{equation}
Here $h_{i,j}({\bf R})$ is the eddy diffusivity that behaves like $H
R^{\zeta_h}$ with $0<\zeta_h<2$.The scaling exponent $\zeta_2$ satisfies
\cite{94Kra} the exact relation $\zeta_2=2-\zeta_h$.

In the inertial range of scales we can disregard the Laplacian operators
in this equation. For deriving the fusion rules (\ref{fusion1}) we
consider the $p$ coalescing points with characteristic separation $r$ and
denote their coordinates by the index $\alpha$ or $\alpha'$. The remaining
$2n-p$ coordinates have characteristic separations $R$ and are denoted by
$\beta$ or $\beta'$. We assemble \cite{95FGLP} the $\hat {\cal B}$
operators in three groups: $\hat {\cal B}_p=\sum_{\alpha>\alpha'} \hat
{\cal B}_{\alpha \alpha'}$, $ \hat {\cal B}_{2n-p}=\sum_{\beta>\beta'}
\hat {\cal B}_{\beta \beta'}$ and $\hat {\cal B}^R=\sum_{\alpha\beta} \hat
{\cal B}_{\alpha \beta}$. The evaluation of the action of the operators in
the first and second groups is $H/r^{\zeta_2}$ and $H/R^{\zeta_2}$
respectively. The evaluation of the action of each term in the third group
is $HR/rR^{\zeta_2}$. However space homogeneity results in a cancellation
of this evaluation in the sum of the terms in this group. The next
order surviving evaluation is again $H/R^{\zeta_2}$. We thus combine the
second and third group into an effective operator $\tilde {\cal B}$. The
equation to consider is
\begin{equation}
\big[\hat {\cal B}_p+\tilde {\cal B}\big]F_{2n} =0 \ . \label{eq}
\end{equation}
When $p=2$ we can find the  solution of (\ref{eq}) as the following
expansion in powers of the small difference $r_{12}$: $A_2\{R\}+
r_{12}^{\zeta_2}C_2\{R\} +
r_{12}^{2\zeta_2}D_2\{R\}+r_{12}^2E_2\{R\}+\dots$, where $A_2$, $C_2$, $D_2$
etc. are some functions of the large separations of the order of $R$. When
we use this solution to compute ${\cal S}_{2n}$ the leading contribution
$A\{R\}$ drops and we find (\ref{fusion1}) for $p=2$. For $p>2$ we
need to distinguish between even and odd $p$'s because of the
special property of passive advection in which ${\cal S}_{2n+1}=0$. The
next even $p$ is $p=4$. For this case we find a solution in the form
\begin{eqnarray}
F_{2n}^c&=&A_4\{R\}+C_4\{R\}\Big[\sum_{\alpha\alpha'=1}^4
r_{\alpha\alpha'}^{\zeta_2}\Big]
\label{F4}\\
&+&D_4\{R\}\big[(r_{12}r_{13})^{\zeta_2}+(r_{12}r_{14})^{\zeta_2}+
(r_{12}r_{23})^{\zeta_2}+\dots\big]
\nonumber\\
&+&F_4^c({\bf r}_1,{\bf r}_2,{\bf r}_3,{\bf r}_4)\Psi_{2n,4}\{R\}+\dots \ ,
\nonumber
\end{eqnarray}
where $F_4^c$ is a contribution of a new type, as it solves the
homogeneous equation (\ref{diffeq}).  Computing ${\cal S}_{2n}$ the first
two terms disappear and in a multiscaling situation the leading
contribution becomes the last. In fact, this is the general rule for any
even order, and is the explicit mechanism for the fusion rules in this
particular case. It arises here from the possibility to split the total
operator $\hat {\cal B}_{2n}$  into the two groups $\hat {\cal B}_p$ and
$\tilde {\cal B}$ such that for $p$ coalescing points $\hat {\cal B}_p$
carries the leading contribution. Since the sum of Laplacians in
(\ref{diffeq}) is also dominated by the sum up to $p$, the crossover scale
$\eta(p,2n,R)$ from inertial range to dissipative behaviour is determined
in this case by a balance between $-\kappa\sum_{\alpha=1}^{p} \nabla^2
_\alpha$ and  $\hat {\cal B}_{p}$. It therefore cannot depend on $n$ or on
$R$: $\eta(p,2n,R)=\eta(p)$. The fusion rules (\ref{Ks})-(\ref{muls})
which were based on the independence of $\eta$ on $R$ follow.

In fact, these results,  and in particular the scaling relations
(\ref{scale}) seem sufficient to determine the exponents $\zeta_n$ in
their entirety.  The necessary steps were detailed in \cite{95FGLP} and will
not be repeated here.

The case of Navier-Stokes turbulence calls for additional considerations.
The fusion rules (\ref{fusion1}), (\ref{fusion2}) were found from first
principles for $p=2$ \cite{95LP-d} and we believe that similar techniques
can be used to establish them for any $p$. Eqs.(\ref{Jpneta})-
(\ref{Jpneta2}) follow, but in the Navier-Stokes case it is possible that
the dissipative scale $\eta(p,n,R)$ is $R$ dependent.  If we assume that
this is not the case the scaling relations obtained above should apply
also to the Navier-Stokes case. The consequences of such an assumption
were discussed in detail in \cite{95LP-d}. To explore another possibility
we follow Kolmogorov's refined similarity hypothesis \cite{62Kol} in
assuming that the conditional average $\nu \langle|\nabla{\bf u}|^2\big|
{\bf w}(0|{\bf R})\rangle \sim  w(0|{\bf R})^3/R$. This assumption means
\begin{equation} \nu J_{2,n}\{R\} \sim S_{n+1}(R)/R \ .
\label{newscale}
\end{equation} Comparing with (\ref{Jpneta}) this can be consistent only if
\begin{equation}
\Big[{\eta(2,n,R)\over \eta(2)}\Big]^{2-\zeta_2}
\sim \Big({R\over L}\Big)
^{\zeta_n-\zeta_{n-1}+\zeta_3-\zeta_2} \ , \label{eta2nr}
\end{equation}
where $\eta(2)$ is by definition the viscous cutoff for the second order
structure function, $\nu S_2(\eta(2))\sim \eta(2)^2 \bar \epsilon$. The
H\"older inequalities guarantee that  $\zeta_n-\zeta_{n-1}$ is a
decreasing function of $n$ in a multiscaling situation.  Accordingly, the
effective dissipative scale of $J_{2,n}$ for two coalescing points
$\eta(2,n,R)$ is much smaller than the viscous cutoff for ${\cal S}_2$,
$\eta(2)$.

Needles to say, with this assumption all our scaling relations change.
For example   consider $\xi(2,n)$ of (\ref{xi}). Instead of (\ref{scale})
we have now
\begin{equation}
\xi(2,n) = \zeta_{n+1} -1   \ . \label{xi2n}
\end{equation}
 Another example is $K_{\epsilon\epsilon}^{(2)}$. we find now
\begin{equation}
\mu(1) = 2-\zeta_6 \ . \label{mu1}
\end{equation}
This result is known as ``the bridge" in the phenomenological theory of
multiscaling turbulence, c.f. \cite{95Nel,Fri}. Notwithstanding the
different scaling relation, the operator algebra that is induced has
``global" and individual scaling exponents as discussed above. The values
of these exponents may be changed due to the $R$ dependence of the
dissipative cutoff as it appears in different models.

In summary, we proposed fusion rules for multiscaling turbulent statistics
that induce an unusual operator algebra. These rules are of two classes.
The first does not involve gradients and is universal for all turbulent
systems with a direct cascade of ``energy" in which there exists a
universal flux equilibrium. The second class involves gradients and these
bring in an explicit dependence on a viscous scale that in general is not
universal.  We explained why in the case of passive scalar advection this
problem may be solved.  Accordingly one can derive an infinite set of non
trivial scaling relations that allow the expression of the scaling exponents of
the correlation functions of gradient fields with the exponents $\zeta_n$
of the structure functions. For Navier-Stokes turbulence the fusion rules
that involve gradients must by supplemented with a theory for the $R$
dependence of the viscous cutoff. We exemplified the influence of a
reasonable assumption about this dependence, but a solid theory of this
dependence based on the Navier-Stokes equations is still a future project.

\noindent
{\bf Acknowledgments}. We thank Mark Nelkin and Uriel Frisch for useful
remarks concerning the inconstancy of the viscous cutoff. This work has
been supported in part by the US-Israel Binational Science Foundation and
the Naftali and Anna Backenroth-Bronicki Fund for Research in Chaos and
Complexity.

\end{document}